\documentclass[aps,prl,twocolumn,groupedaddress,showpacs]{revtex4}
\usepackage{graphics,epsfig}

\begin{document}


\title{A High Sensitivity Search for $\bar{\nu}_{e}$'s from the Sun and Other Sources at KamLAND}



\author{
K.~Eguchi,$^1$
S.~Enomoto,$^1$
K.~Furuno,$^1$
H.~Ikeda,$^1$ 
K.~Ikeda,$^1$
K.~Inoue,$^1$
K.~Ishihara,$^{1*}$
T.~Iwamoto,$^{1\dagger}$
T.~Kawashima,$^1$
Y.~Kishimoto,$^1$
M.~Koga,$^1$
Y.~Koseki,$^1$
T.~Maeda,$^1$
T.~Mitsui,$^1$
M.~Motoki,$^1$
K.~Nakajima,$^1$
H.~Ogawa,$^1$
K.~Owada,$^1$
F.~Piquemal,$^1$
I.~Shimizu,$^1$
J.~Shirai,$^1$
F.~Suekane,$^1$
A.~Suzuki,$^1$
K.~Tada,$^1$
O.~Tajima,$^1$
K.~Tamae,$^1$
H.~Watanabe,$^1$
J.~Busenitz,$^2$
Z.~Djurcic,$^2$
K.~McKinny,$^2$
D-M.~Mei,$^2$
A.~Piepke,$^2$
E.~Yakushev,$^2$
B.E.~Berger,$^3$
Y.D.~Chan,$^3$
M.P.~Decowski,$^3$
D.A.~Dwyer,$^3$
S.J.~Freedman,$^3$
Y.~Fu,$^3$
B.K.~Fujikawa,$^3$
J.~Goldman,$^3$
K.M.~Heeger,$^3$
K.T.~Lesko,$^3$
K.-B.~Luk,$^3$
H.~Murayama,$^3$
D.R.~Nygren,$^3$
C.E.~Okada,$^3$
A.W.P.~Poon,$^3$
H.M.~Steiner,$^3$
L.A.~Winslow,$^3$
G.A.~Horton-Smith,$^4$
C.~Mauger,$^4$
R.D.~McKeown,$^4$
B.~Tipton,$^4$
P.~Vogel,$^4$
C.E.~Lane,$^5$
T.~Miletic,$^5$
P.W.~Gorham,$^6$
G.~Guillian,$^6$
J.G.~Learned,$^6$
J.~Maricic,$^6$
S.~Matsuno,$^6$
S.~Pakvasa,$^6$
S.~Dazeley,$^7$
S.~Hatakeyama,$^7$
R.~Svoboda,$^7$
B.D.~Dieterle,$^{8}$
M.~DiMauro,$^{8}$
J.~Detwiler,$^{9}$
G.~Gratta,$^{9}$
K.~Ishii,$^{9}$
N.~Tolich,$^{9}$
Y.~Uchida,$^{9}$
M.~Batygov,$^{10}$
W.~Bugg,$^{10}$
Y.~Efremenko,$^{10}$
Y.~Kamyshkov,$^{10}$
A.~Kozlov,$^{10}$
Y.~Nakamura,$^{10}$
C.R.~Gould,$^{11}$
H.J.~Karwowski,$^{11}$
D.M.~Markoff,$^{11}$
J.A.~Messimore,$^{11}$
K.~Nakamura,$^{11}$
R.M.~Rohm,$^{11}$
W.~Tornow,$^{11}$
A.R.~Young,$^{11}$
M-J.~Chen,$^{12}$
and Y-F.~Wang$^{12}$\\
(KamLAND Collaboration)
}
\affiliation{
$^1$ Research Center for Neutrino Science, Tohoku University, Sendai
980-8578, Japan \\
$^2$ Department of Physics and Astronomy, University of Alabama, Tuscaloosa,
Alabama 35487, USA \\
$^3$ Physics Department, University of California at Berkeley and Lawrence Berkeley National
Laboratory, Berkeley, California 94720, USA \\
$^4$ W.~K.~Kellogg Radiation Laboratory, California Institute of
Technology, Pasadena, California 91125, USA \\
$^5$ Physics Department, Drexel University, Philadelphia, Pennsylvania
19104, USA \\
$^6$ Department of Physics and Astronomy,
University of Hawaii at Manoa, Honolulu, Hawaii 96822, USA \\
$^7$ Department of Physics and Astronomy, Louisiana State University, Baton Rouge, Louisiana
70803, USA \\
$^{8}$ Physics Department, University of New Mexico, Albuquerque,
New Mexico 87131, USA \\
$^{9}$ Physics Department, Stanford University, Stanford, California
94305, USA \\
$^{10}$ Department of Physics and Astronomy, University of Tennessee, Knoxville, Tennessee
37996, USA \\
$^{11}$ Triangle Universities Nuclear Laboratory, Durham, North Carolina 27708, USA
and \\
Physics Departments at Duke University, 
North Carolina State University, and 
 the University of North Carolina at Chapel Hill \\
$^{12}$ Institute of High Energy Physics, Beijing 100039, People's Republic of China
}%


\date{\today}

\begin{abstract}
Data corresponding to a KamLAND detector exposure of 0.28 kton-year has been used to search for $\bar{\nu}_e$'s in 
the energy range 8.3~MeV~$<$ E$_{\bar{\nu}_e}$ $<$~14.8 MeV.  No 
candidates were found for an expected background of $1.1{\pm}0.4$ events.
This result can be used to obtain a limit on $\bar{\nu}_{e}$ fluxes of any
origin.
Assuming that all $\bar{\nu}_e$ flux has its origin in the Sun and has the characteristic $^8$B solar $\nu_e$
energy spectrum, we obtain an upper limit  
of 3.7~$\times$~10$^2$~cm$^{-2}$ s$^{-1}$ (90\% C.L.) on the $\bar{\nu} _e$ flux. We interpret this limit, corresponding to $2.8{\times}10^{-4}$ of the Standard Solar Model $^8$B $\nu_e$ flux, in the framework of spin--flavor precession and neutrino decay models.
\end{abstract}


\maketitle



Of the many mechanisms that have been suggested to explain the
solar neutrino problem~\cite{solarprob}, neutrino oscillations are
strongly favored by the data.  
Assuming CPT invariance, the recent observation of reactor
$\bar{\nu}_e$ disappearance by the Kamioka Liquid Scintillator Anti-Neutrino Detector (KamLAND)~\cite{Eguchi}, combined with
direct measurements of the solar neutrino flux~\cite{solarexpt}, indicates
that the oscillation parameters lie in the Mikheyev-Smirnov-Wolfenstein (MSW)~\cite{msw} Large Mixing Angle (LMA) region~\cite{lma}. 
However, the limited precision of current
measurements still allows for the possibility that other mechanisms play
a sub-dominant role. Since further study of the nature of neutrinos and the properties of the Sun is vital, we report in this {\it Letter} on a
search for solar $\bar{\nu}_e$'s.

There are several conceivable
mechanisms which would lead to a $\bar{\nu}_e$ component in the solar
flux incident on Earth. Electron neutrinos with a non-zero transition magnetic moment can evolve into $\bar{\nu}_{\mu}$'s or $\bar{\nu}_{\tau}$'s
while propagating
through intense magnetic fields in the solar core. These neutrinos can, in turn, evolve into $\overline{\nu}_{e}$'s via 
flavor oscillations. There is also neutrino
decay, in which a heavy neutrino mass eigenstate decays into a lighter
anti-neutrino mass eigenstate~\cite{bbdecay,decay}.

The analysis presented in this {\it Letter} concerns a
search for $\bar{\nu}_{e}$'s regardless of origin. Possible non-solar sources
of $\bar{\nu}_{e}$'s at KamLAND include Weakly Interacting Massive Particle
(WIMP) annihilation in the Sun and Earth~\cite{wimp} and relic supernova neutrinos~\cite{relic,supernu},
either of which could contribute to a continuous $\bar{\nu}_{e}$ flux. The
event rates from these~\cite{estimate} and other non-solar sources are
expected to be small, however, and we choose to focus on
models that predict a flux of $\bar{\nu}_{e}$'s descendant from solar
neutrinos.

KamLAND was designed to study the flux of reactor $\bar{\nu}_e$'s.  While
the reactor $\bar{\nu}_{e}$ flux spectrum has an endpoint of $\sim$~8.5 MeV, the $^8$B
solar neutrino flux spectrum extends well beyond this energy to $\sim$~15
MeV.  As a result, KamLAND data may be used to search for
$\bar{\nu}_e$'s in the solar neutrino flux over an energy range largely free of
reactor $\bar{\nu}_e$ events.


The detector consisted of a thin
plastic--walled balloon, 13\,m in diameter, filled with about 1~kton of
liquid scintillator (7.6$\times$10$^{31}$ free protons).  The balloon was
surrounded by an 18--meter-diameter stainless steel sphere
instrumented with 1325 17--inch and 554 20--inch Hamamatsu photomultiplier tubes
(PMTs), which provided 34\% photo-coverage.  For the search presented here, only the data from
the 17--inch PMTs were analyzed, lowering the photo-coverage to 22\%.  The space between the stainless
steel sphere and the balloon contained a mixture of dodecane and isoparaffin
oils to act as a buffer against external backgrounds.
The stainless steel sphere and its contents (hereafter referred to as the inner detector (ID)) was itself contained within a cylindrical water
Cerenkov outer detector (OD) equipped with 225 20--inch PMTs.
The OD was used to tag events due to cosmic ray
induced particles.  The entire detector was shielded by a rock
overburden of about 1000\,m (2700\,m.w.e.), which reduced the cosmic muon flux
by a factor of 10$^{5}$ relative to that at the surface.


    Electron anti--neutrinos were detected via 
    the inverse $\beta$--decay reaction
    \begin{eqnarray}
    \bar{\nu}_{e} + p & \rightarrow & e^{+} + n, \label{ibd1}
    \end{eqnarray} 
    \noindent consisting of a prompt energy
    deposit from the positron and two annihilation $\gamma$'s followed
    $\sim$210\,$\mu$s later by neutron capture on hydrogen, producing
    a 2.2\,MeV $\gamma$.  For the energy range of our search, this
    reaction, in which the final state neutron is free, can 
    occur only on the free proton of the hydrogen nucleus.
The $\bar{\nu}_{e}$ energy was deduced from the prompt
energy E$_{prompt}$ using the 
relationship E$_{\bar{\nu}_{e}}$ = E$_{prompt}$~+E$_{recoil}$+~0.8 MeV, where
the small quantity E$_{recoil}$ refers to the neutron kinetic energy in the
final state and was neglected.

\begin{figure}
\epsfxsize=0.5\textwidth\epsfbox{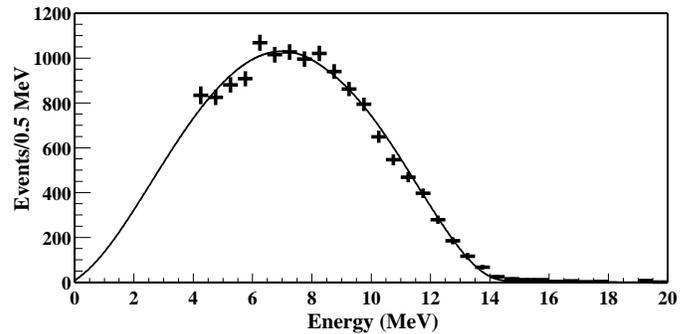}
\caption{\label{fig:pvsd1} Prompt energy spectrum of
$^{12}$B decay. Points are KamLAND data and the curve is the expected
$\beta$ decay spectrum convolved with the detector response.}
\end{figure}

Event reconstruction for high energy inverse $\beta$--decay 
events in this analysis was similar to that described in~\cite{Eguchi} and  
was found to be accurate to within 2\% from comparison with
the observed energy distribution of the $\beta$ decay of cosmogenically
produced $^{12}$B and $^{12}$N (Figure~\ref{fig:pvsd1}).
The measured energy resolution of KamLAND for this data set was 7.5\%/$\sqrt{E_{prompt}(\mbox{MeV})}$. 
Events with $8.3~\mbox{  MeV}<E_{\bar{\nu}_{e}}<14.8~\mbox{  MeV}$,
followed \mbox{0.5\,$\mu$s~$-$\,660~$\mu$s} later by a delayed event
depositing between $1.8\,\mathrm{MeV}$ and $2.6\,\mathrm{MeV}$ of
energy, were selected.  The distance between the prompt and
delayed vertices was constrained to be less than 160~cm and both vertices were
required
to be within 550\,cm of the detector center in order to suppress backgrounds
due to natural radioactivity and muon spallation. Backgrounds were
further reduced by using ID PMTs to reconstruct a muon track for all events
containing OD data. Anti-neutrino candidates associated with
detected muons were discarded if they occurred within 2~s after un-reconstructed muons, within 
2~s after muons depositing at least 3
GeV, or within 2~s and less than 3~m from a
reconstructed muon track. Spallation neutrons associated with tagged muons
were also removed and did not contribute to the background.

\begin{figure}
\epsfxsize=0.5\textwidth\epsfbox{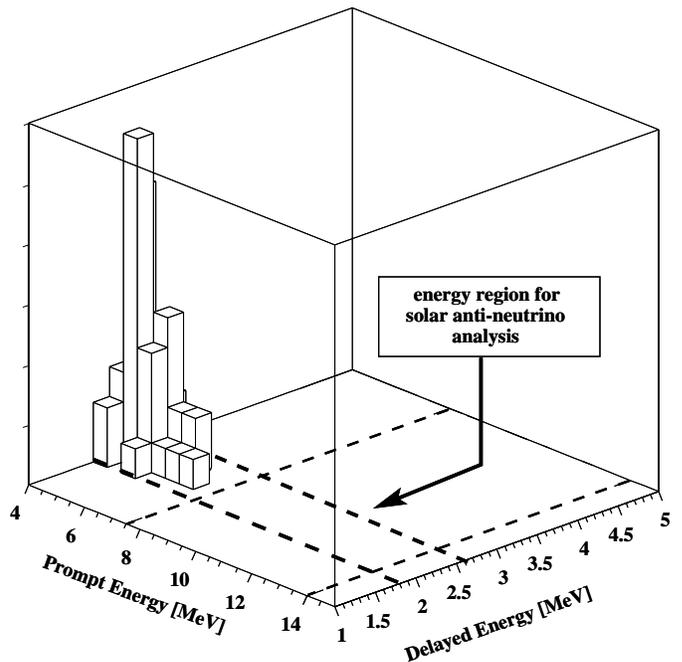}
\caption{\label{fig:pvsd2} Energy distribution of the final event candidates.
The tail from reactor $\bar{\nu}_{e}$ events is visible below 8~MeV.}
\end{figure}

Figure~\ref{fig:pvsd2} shows the delayed versus prompt energy distribution for events after all selection 
cuts, except those on the prompt and delayed energies themselves. Taking into account the 12\%
deadtime associated with muon rejection, the total sample livetime, corresponding to the period March 4 - December 1, 2002, was 185.5 days.

The detection efficiency for inverse $\beta$--decay events was estimated from Monte
Carlo simulation and calibration data to be \mbox{84.2~$\pm$~1.5\%}.  The main contributions to the detection 
inefficiency were the cuts on the distance between the prompt and delayed vertices 
\mbox{(89.8~$\pm$~1.6\%)}, the time between the prompt and delayed vertices \mbox{(95.3~$\pm$~0.3\%)}, neutron capture on protons \mbox{(99.5\%)},
and the energy of the delayed event {(98.9~$\pm$~0.1~\%)}.
The efficiency of the vertex-separation cut was determined by a Monte Carlo
simulation checked against AmBe neutron source data. The neutron
capture time distribution with mean $210 {\pm} 5\,\mu\mbox{s}$ and the delayed
energy cut efficiency were measured using both AmBe neutron and spallation
neutron data.

No events were observed in the signal region over the 0.28 kton-year sample. A
separate analysis of the KamLAND data was carried
out as a cross-check using a subset of the 0.28 kton-year sample presented here. 
The results of both analyses were consistent.


The energy spectrum of reactor $\bar{\nu}_e$'s extends to about 8.5\,MeV and
may have constituted a small background in the solar $\bar{\nu}_e$ energy region
due to the detector's finite energy resolution.
The number of background events was estimated to be \mbox{0.2~$\pm$~0.2} and
uncertainties in the detector energy scale and neutrino oscillation parameters 
were the dominant sources of error. The background from atmospheric
neutrinos was estimated, using the Barr-Gaisser-Stanev flux~\cite{bgs}, to be
0.001 events.

Cosmic ray muons interact in and near the detector producing spallation
neutrons with an energy spectrum extending up to several hundred MeV.  These
neutrons constituted a source of background for
the inverse $\beta$--decay measurement as the prompt deposit of the neutron's kinetic energy followed $\sim$210\,$\mu$s 
later by the capture of the thermal neutron was indistinguishable from the
inverse $\beta$--decay event signature. As outlined below, we estimated the
spallation neutron contribution to the background using a sample of
neutrons selected from the data.

Spallation neutron cuts were the same as for $\bar{\nu}_e$ candidates except
that the fiducial volume cut was dropped and the muon-related cuts
were replaced by the requirement that at least 5 PMTs in the OD fired.
The radial distribution of the remaining candidates
was fitted in order to obtain a smooth extrapolation of
the fast neutrons into the fiducial volume. The resulting fitted function was
integrated inside the volume to estimate the expected number of
fast neutron events $N _{fn}$ meeting the selection criteria.  
We used this quantity to estimate the two components of
the fast neutron background by multiplying $N _{fn}$ by a factor of 0.11,
determined from Monte Carlo calculations, to obtain the contribution from fast neutrons due to muons 
passing through the rock near the detector and by scaling $N _{fn}$ by the OD
detector inefficiency to obtain the contribution from fast neutrons produced by muons passing through the OD but missing the ID.  Summed, these two components
contributed \mbox{0.3~$\pm$~0.2 events} to the background.

We estimated the background due to accidental coincidences using data events
falling within an off-time delayed coincidence window of 1--10~s. Two hundred
and seventeen such coincidences were found, corresponding to a background contribution of
0.02 events after normalization to the width of the $\bar{\nu}_e$ delayed coincidence window.

The residual backgrounds from cosmogenic $^8$He ($t_{1/2}=0.12$~s) and $^9$Li
($t_{1/2}=0.18$~s) decays were estimated by determining the total number of
these events in the data sample and extrapolating into the $\bar{\nu}_e$
signal region using known decay times and vertex distributions.
Above 8.3~MeV, the $^9$Li contribution dominated and, accordingly, analyses
in that energy region dealt exclusively with $^9$Li. The residual
contribution to the background was calculated to be \mbox{0.6~$\pm$~0.2}
events.

Table \ref{bg} summarizes the background estimates for this data set.

\begin{table}[h]
\begin{ruledtabular}
\begin{tabular}{ l c }
Background Source & Expected Events \\ \hline
Reactor $\bar{\nu} _e$ & 0.2$~\pm$~0.2 \\ 
Atmospheric neutrinos   & 0.001  \\
Fast neutrons ($N_{fn}$) & 0.3~$\pm$~0.2  \\
Accidental coincidences & 0.02 \\
$^8$He \& $^9$Li         & 0.6~$\pm$~0.2 \\ \hline
Total                   & 1.1~$\pm$~0.4 \\
\end{tabular}
\end{ruledtabular}
 \caption[Backgrounds for inverse $\beta$--decay]
 {Estimated backgrounds for the inverse $\beta$--decay signal in the energy range of 8.3 MeV~$<$~$E_{\bar{\nu}_{e}}$~$<$~14.8 MeV for 185.5 live-days.
  \label{bg}}
\end{table}


The $\bar{\nu}_e$ flux integrated over 
the energy range \mbox{8.3--~14.8~MeV} is obtained from:

\begin{equation}
\Phi_{\bar{\nu}_e} = \frac{N_{signal}}{  \bar{\sigma} \times \bar{\epsilon} \times T \times \rho_{p} \times f_{v} },
\label{fluxcalc}
\end{equation}

\noindent where $N_{signal}$ is the number of 
detected $\bar{\nu} _e$'s, \mbox{$\bar{\sigma}$~=~6.9~$\times$~10$^{-42}$\,cm$^2$} 
and {\mbox{$\bar{\epsilon}$~=~0.84} are the  cross section~\cite{invbxsec}
 and detection efficiency, respectively, averaged over energy, \mbox{T~=~1.6~$\times$~10$^{7}$\,s} 
is the livetime, and \mbox{$\rho_p{\times}f_{v}$~=~4.6~$\times$~10$^{31}$} is
the number of target protons in the fiducial volume $f_{v}$ (radius 550~cm).
For calculating the average cross section and
detection efficiency, the shape of the Standard Solar Model $^8$B flux
without
oscillations~\cite{ortiz} was used.

Systematic uncertainties in the quantities in Equation~\ref{fluxcalc} are tabulated in
Table~\ref{syst}.
The systematic uncertainty in the number of target protons \mbox{($\rho_p{\times}f_{v}$)} was obtained by adding in quadrature the 2.1\% uncertainty in the amount of
scintillator in the balloon and the estimated 3.7\% uncertainty in the
fiducial volume. This latter estimate is based on
the difference between the measured number of spallation products
in the fiducial volume and the expected number assuming that the
spallation products were uniformly distributed. The contribution
from the energy threshold was calculated using the uncertainties
in the energy scale (2\%) and the slope of the neutrino flux
at the threshold of 8.3~MeV.

\begin{table}[h]
\begin{ruledtabular}
\begin{tabular}{ l c }
Quantity & Systematic Uncertainty (\%) \\ \hline
Detection efficiency ($\bar{\epsilon}$)  & 1.6 \\
Cross section ($\bar{\sigma}$) & 0.2 \\
Number of target protons & 4.3 \\
Energy threshold & 4.3 \\
Livetime (T)                   & 0.07 \\ \hline
Total                          & 6.3 \\
\end{tabular}
\end{ruledtabular}
 \caption[Total uncertainty]
 {Systematic uncertainties in quantities used to determine the flux of solar $\bar{\nu}_e$.  
  \label{syst}}
\end{table}

We derived an upper limit on $\Phi_{\bar{\nu}_e}$ using the Feldman-Cousins
unified approach~\cite{feldcous} supplemented with Bayesian modifications to account for the errors on
nuisance~\cite{pdg} and background parameters~\cite{coushlnd,conrad,hill}. For no observed events, the upper limit of the
$\bar{\nu}_e$ flux was \mbox{3.7~$\times$~10$^2$~$\mathrm{cm}^{-2} \mathrm{s}^{-1}$} at 90\% CL. Using the
prescription described in \cite{feldcous}, the sensitivity of this
measurement was \mbox{$7.9\times10^{2}\:\mathrm{cm^{-2}s^{-1}}$}\ (90\%
C.L.). Normalizing to the solar $\mathrm{^{8}B}$\ $\nu_{e}$\
flux~\cite{bahcall} in the
analysis energy window ($\mathrm{8.3\:MeV<E_{\nu_{e}}<14.8\:MeV}$, containing
29.5\% of the total flux  of
$5.05{\times}10^{6}\:(1.00^{+0.20}_{-0.16})\:\mathrm{cm^{-2}s^{-1}}$~\cite{bahcall}), 
this flux limit corresponds to an upper limit on the neutrino conversion probability 
of $2.8{\times}10^{-4}$ at the 90\% C.L. and represents a factor of 30
improvement over the best previous measurement~\cite{sknuebar}.

We have assumed a non-oscillatory solar $\bar{\nu_{e}}$ flux up to now
in order to retain as much generality as possible but,
in the following, we have interpreted the KamLAND upper limit on the solar
$\overline{\nu}_{e}$ flux
in the framework of two models: spin-flavor precession combined with 
neutrino oscillations and neutrino decay.  

  Assuming that the solution to the solar neutrino problem lies within
the LMA region of parameter space and that the MSW effect is a dominant mechanism affecting the
solar neutrino flux, we followed the treatment of~\cite{sfp,akhmedov} (taking the value of 34 degrees for the mixing angle) and obtained the following limit on the
product of the neutrino transition magnetic moment $\mu$ and the transverse
component of the magnetic field $B _T$ in the Sun at a radius of $0.05R_s$:
   
\begin{eqnarray}
\frac{\mu}{10^{-12}\mu_B}\frac{B_T(0.05R_s)}{10~\mathrm{kG}} < 
1.3 \times 10 ^3 
\end{eqnarray}

\noindent The current best limit on the neutrino magnetic moment is from the
MUNU experiment~\cite{munu}: $\mu_{\overline{\nu}_{e}}<1.0{\times}10^{-10}\mbox{ }\mu_{B}$ (90\% C.L.).

Similarly, for quasi-degenerate neutrino masses, we were able to constrain the
lifetime~\cite{bbdecay,decay} for $\nu_2$, the heavier neutrino, to
\mbox{$\tau _2/m_2$~$>$~0.067~s/eV}. 
If the neutrino mass spectrum is hierarchical,
the limit is weaker and for $m_2$  of about 0.01
eV~($\sim \sqrt{\Delta m_{12} ^2}$), $\tau_2$~$>$~11~$\mu$s. 
Our limit, obtained using the appropriate decay branching ratio for final states 
containing a $\bar{\nu} _e$, represents
an improvement over the current bound of $\tau/m$~$>$~$10^{-4}$~s/eV
suggested 
in~\cite{bbdecay}.
%


To summarize, we have described a search for $\bar{\nu}_e$'s in the energy
range \mbox{(8.3~MeV~$<$~$E_{\bar{\nu}_{e}}$~$<$~14.8 MeV)}
with KamLAND. The KamLAND detector's source-independent
sensitivity allows for the measurement of $\bar{\nu}_{e}$ fluxes
independent of origin. No events were found in the 185.5 live-day data set,
allowing for an upper limit to be set on the flux from any source producing
$\bar{\nu}_{e}$'s in the appropriate energy range. We have obtained a flux
limit of $\Phi_{\bar{\nu}_e} $~$<$~3.7~$\times$~10$^2$~cm$^{-2}$ s$^{-1}$
(90\% C.L.), assuming a solar origin and an un-oscillated $^8$B neutrino
energy spectrum. This limit has been used to constrain models of neutrino
spin--flavor precession and neutrino decay.


The KamLAND experiment is supported by the COE program of the Japanese Ministry of
Education, Culture, Sports, Science, and Technology and the United States Department
of Energy.  We are grateful to the Kamioka Mining and Smelting Company 
for providing services at the experimental site.\\

{\small
{\it
\noindent $^{*}$ Present address: Institute for Cosmic Ray Research, University of Tokyo\\
$^{\dagger}$ Present address: International Center for Elementary Particle Physics, University of Tokyo}
}


\begin{thebibliography}{99}
\bibitem{estimate} For a WIMP mass of 100 GeV, on the order of 0.06 $\bar{\nu} _e$'s
  interactions per kton-year are expected at KamLAND; while the expectation of the event rate
  from relic  $\bar{\nu} _e$'s is limited to approximately 0.1 events per
  kton-year.
\end{thebibliography}

\end{document}